\documentstyle[12pt]{article}
\newcommand{\sect}[1]{\setcounter{equation}{0}\section{#1}}
\renewcommand{\theequation}{\thesection.\arabic{equation}} \begin{document}
\def\bq{\begin{equation}}
\def\eq{\end{equation}}
\begin{flushright}
{April,1998}\\
LPTENS-98/11\\
\end{flushright}

\begin{center}
{\large\bf
{LEVEL SPACING
OF RANDOM MATRICES IN AN EXTERNAL SOURCE }} \end{center}

\begin{center}
{\bf{ E. Br\'ezin$^{a)}$ and S. Hikami$^{b)}$}} \end{center}
\vskip 2mm
\begin{center}
$^{a)}$ Laboratoire de Physique Th\'eorique, Ecole Normale Sup\'erieure\\
24 rue Lhomond 75231, Paris Cedex 05,
France
\footnote{ Unit\'e propre 701 du Centre national de la recherche
scientifique, Associ\'ee \`a l'Ecole Normale Sup\'erieure et \`a
l'Universit\'e de Paris-Sud}\\
$^{b)}$ Department of Pure and Applied Sciences, University of Tokyo\\
Meguro-ku, Komaba, Tokyo 153, Japan\\ \end{center} \vskip 3mm
\begin{abstract}
In an earlier work we had considered a Gaussian ensemble of random matrices
in the presence of a given external matrix source.
The measure is no longer unitary invariant and the usual techniques based
on orthogonal polynomials, or on the Coulomb gas representation, are not
available.
Nevertheless the n-point correlation functions are still given in terms of
the determinant of a kernel, known through an explicit integral
representation. This
kernel is no longer symmetric though and is not readily accessible to
standard methods. In particular finding the level spacing probability is
always a delicate
problem in Fredholm theory, and we have to reconsider the problem within
our model. We find a new class of universality for the level spacing
distribution when the
spectrum of the source is ajusted to produce a vanishing gap in the density
of the state. The problem is solved through coupled non-linear differential
equations,
which turn out to form a Hamiltonian system. As a result we find that the
level spacing probability $p(s)$ behaves like $\exp[ - C s^{{8\over{3}}}]$
for
large spacing
$s$; this is consistent with the asymptotic behavior $\exp[ - C s^{2 \beta
+ 2}]$, whenever the density of state behaves near the edge as
$\rho(\lambda)\sim
\lambda^{\beta}$. \end{abstract}
\newpage

\sect{INTRODUCTION}

The level spacing distribution $p(s)$, first discussed by Wigner
\cite{Wigner} for nuclear energy levels, has been studied extensively in
random matrix theory \cite{Dyson} and the universality of $p(s)$ has been
discussed for many cases, including the the distribution of zeros in
Riemann zeta function\cite{Mehta}. The calculation of this level spacing
distribution is always much more delicate than that of the n-point
correlation functions. Indeed when two neighboring levels are separated by
some interval $s$, it implies that all the other eigenvalues are outside of
this interval, and consequently it involves all the correlation functions
of those eigenvalues. In the simplest Gaussian ensemble it took many years
of development of the theory of Fredholm determinants, Dyson's inverse
scattering approach, tau-functions , before
this problem was understood. For the Airy kernel which appears for the
spacing in the vicinity of the edge of Wigner's semi-circle, Tracy and
Widom developed a
technique through coupled non-linear differential equations, which we have
generalized here and applied to our previous work on random matrices in a
matrix source\cite{BH4}.

The problem we had studied concerns a Hamiltonian $H=H_0+V$, in which $H_0$
is an $N$ by $N$ non-random matrix with a known spectrum, but $V$ is a
random Gaussian potential. The probability measure for $H$, beeing then a
Gaussian in $H-H_0$, is not unitary invariant. Consequently the standard
approach,
which consists of tracing out the unitary degrees of freedom, and then
using orthogonal polynomials, or a Coulomb gas representation, is not
available.
However we have obtained exact formulae, i.e. valid for matrices of finite
size, for the n-level correlation functions with the help of the
Itzykson-Zuber formula \cite{Itzykson}. It is remarkable that these n-point
functions are still given by a determinant of an n by n matrix whose
elements are
given by a kernel $K(\lambda,\mu)$, a well-known fact \cite{Mehta} when the
orthogonal polynomials are available. However in our problem, this kernel
is no longer symmetric, but we know an exact integral representation for
finite $N$.

If we had considered the level spacing distribution for those ensembles
with a source, around a regular point of the spectrum, the usual proofs of
universality would apply and the end result, in the appropriate scaling
limit, would be identical to the Wigner ensemble level spacing. However
near singularities of the spectrum new universality classes appear. For
instance near the edge of the Wigner semi-circle, in a region of size
$N^{-2/3}$, a new kernel expressed in terms of Airy functions controls the
correlation functions; Tracy and Widom have succeeded to find the level
spacing
for this new kernel.

In the source problem, we have shown earlier that we can have gaps in the
spectrum of $H$, when the randomness of $V$ is not strong enough to bridge
the gaps between the eigenstates of $H_0$. If we tune the randomness in
order to reach the limiting situation in which a gap closes, a new
universal singularity appears at which the density of eigenvalues
$\rho(\lambda)$ vanishes like $\vert\lambda\vert^{1/3}$. In this paper we
obtain the level
spacing distribution by an application of Fredholm theory from which we
derive non-linear differential equations, which remarkably form a
Hamiltonian system. This generalizes earlier work, and makes it clear that
the technique is general. For the Wigner case, with the sine-kernel
obtained by Dyson,
\bq\label{1.1}
K(x,y) = {\sin \pi (x - y)\over{\pi (x - y)}} \qquad, \eq Jimbo et al
\cite{Jimbo} had obtained long ago a closed equation for E(s), the
probability that the interval
$(-s/2, s/2)$ is empty.
Tracy and Widom \cite{Tracy1} had considered the Airy kernel \bq\label{1.2}
K(x,y) = {A_i(x) {A_i}^{\prime}(y) - {A_i}^{\prime}(x) A_i(y)\over{x - y}}
\eq (where $A_i(x)$ is an Airy function which satisfies
${A_i}^{\prime\prime}(x) = x A_i(x)$), and obtained $E(s)$ for the
semi-infinite interval $(s,\infty)$ , when $s$ is near a singular edge of
the semi-circle. The method developed in this paper is easily applicable to
those two cases, which are briefly reviewed in the Appendices A and B.

In all those earlier cases, as well as for our source problem, the
probability $E(s)$ of emptiness of the interval (-s/2, s/2), is a Fredholm
determinant,
\bq\label{1.3}
E(s) = \sum_{n=0}^{\infty} {(-1)^{n}\over{n!}} \int_a^b \cdots \int_a^b
\Pi_{k=1}^{n} d x_k {\rm det}[ K(x_i,x_j)]_{i,j=1,...,n} \eq if we choose
$(a=-s/2, b=+s/2)$ for the interval (a,b). The sine and Airy kernels are
symmetric : $K(x,y) = K(y,x)$.
For the source problem $K(x,y)$ is no longer symmetric under exchange of x
and y, although its square under convolution yields back the kernel itself
as for the simpler cases\cite {BH4}. Starting with Fredholm theory for
$E(s)$ and extending the analysis of Tracy and Widom
\cite{Tracy1,Tracy2,Tracy3}, we derive for our problem a Hamiltonian
system, leading to coupled non-linear differential equations. From those
equations one can determine the function $E(s)$, and in particular its
asymptotic expansion for large $s$, which is interesting. For the sine
kernel, it behaves as $E(s) \sim \exp( - {\pi^2\over{8}}s^2)$; for the Airy
kernel, it becomes
$E(s)\sim \exp(- {1\over{96}} s^3)$.
In our previous work\cite{BH4}, we had applied a Pad\'e analysis to a
function $R(t)$ related to
$E(s)$,
\bq\label{1.4}
E(s) = \exp[ \int_0^{\tilde s} R(\tilde s) d\tilde s] \eq where the
variable $\tilde s$ is defined as
\bq\label{1.5}
\tilde s = \int_{-s/2}^{s/2} \rho(x)
dx
\eq
We had made the ansatz in our previous paper \cite{BH4} that $R(\tilde s)$
behaves like
$\tilde s$ in the large $\tilde s$ limit, namely that $E(s)$ is Gaussian in
terms of $\tilde s$.
More generally the ansatz was that for a density of state behaving as
$\rho(x) \sim x^{\beta}$, then
\bq\label{1.6}
E(s) \sim \exp[ - C s^{2\beta + 2}]
\eq
Indeed this ansatz agrees with the results for the sine-kernel, for which
$\beta = 0$, and for Airy kernel $\beta = 1/2$. It also agrees with our gap
closing singularity for which the density of state gives an exponent $\beta
= 1/3$. The result of the present analysis confirms this conjecture since
it gives $E(s) \sim \exp[ - C s^{8/3}]$.

\sect{The kernel at the closure of a gap}

We consider an $N \times N$ Hamiltonian matrix $H = H_0 + V$, where $H_0$
is a given
non-random Hermitian matrix, and
$V$ is a random Gaussian
Hermitian matrix \cite{BH4,BHZ1,BH1,BH2,BH3}. The probability distribution
$P(H)$ is thus given by \begin{eqnarray}\label{2.1} P(H) &=& {1\over{Z}}e^{
- {N\over{2}} {\rm Tr} V^2 }\nonumber\\ &=& {1\over{Z^{'}}}e^{-
{N\over{2}}{\rm Tr} ( H^{2}- 2 H_0 H)} \end{eqnarray} For the n-point
correlation functions, defined as \bq\label{2.2} R_{n}(\lambda_1,\lambda_2,
\cdots, \lambda_n) = <{1\over{N}} {\rm Tr}\delta(\lambda_1 - H)
{1\over{N}}{\rm Tr}
\delta(\lambda_2 - H ) \cdots{1\over{N}} {\rm Tr}\delta(\lambda_n - H ) >
\eq we have derived earlier\cite{BH2,BH3} the expression

\bq\label{2.3}
R_2(\lambda_1, \lambda_2) = K_N(\lambda_1, \lambda_1) K_N(\lambda_2,
\lambda_2) -
K_N(\lambda_1, \lambda_2) K_N(\lambda_2, \lambda_1) \eq with the kernel
\bq\label{2.4}
K_N(\lambda_1, \lambda_2) = (-1)^{N-1}\int {dt\over{2 \pi}}\oint {du\over{2
\pi i}}\prod_{\gamma =
1}^{N} ( {
a_\gamma - i t\over{ u - a_\gamma}}) {1\over{u - i t}} e^{- {N\over{2}}u^2
- {N\over{2}} t^2 - N i
t \lambda_1 + N u \lambda_2}
\eq
Similarly the n-point functions are given in terms of the determinant of
the $n\times n$ matrix
whose elements are given by this same kernel $ K_N(\lambda_i,\lambda_j)$
\cite{BH3,Guhr}.
In \cite{BH1}, this kernel $K_N(\lambda_1, \lambda_2)$ was considered in
the large N limit, for fixed $N(\lambda_1 - \lambda_2)$. In this limit one
can evaluate the integrals (\ref{2.4}) by the saddle-point method. The
result was found to be, up to a phase factor that we omit here,
\bq\label{2.5} K_N(\lambda_1, \lambda_2) = - {1\over{\pi y}} {\rm sin} [\pi
y \rho(\lambda_1)] \eq
where
$y = N(\lambda_1 - \lambda_2)$.
Apart from the scale dependence provided by the density of state $\rho$,
the n-point correlation
functions have thus
a universal scaling limit, i.e. independent of the deterministic part $H_0$
of the
random Hamiltonian.

In order to generate a tunable gap we consider the simple case for which
the eigenvalues of $H_0$ are $\pm a$, each value being N/2 times
degenerate. When the randomness of $V$ is small, the average density of
eigenvalues is made
of two disjoint segments located around the points $\pm a$. When the
randomness increases one reaches
a critical point at which the gap closes. When that happens in the vicinity
of the origin of size $N^{-3/4}$
we find a new class of universality for the density of states and for the
n-point functions\cite{BH4}. Then the kernel (\ref{2.4}) becomes

\bq\label{2.6}
K_N(\lambda_1, \lambda_2) = (-1)^{{N\over{2}} + 1} \int {dt\over{2
\pi}}\oint {du\over{2 \pi i}}({a^2 + t^2\over{u^2 - a^2}})^{{N\over{2}}}
{1\over{u - i t}} e^{- {N\over{2}}u^2 - {N\over{2}} t^2 - N i t \lambda_1 +
N u
\lambda_2}
\eq
From this expression,
we obtain the density of state $\rho(\lambda)= K_N(\lambda,\lambda)$. The
derivative
of $\rho(\lambda)$ with respect to
$\lambda $ eliminates the factor $u - i t$ in the denominator of
(\ref{2.4}), and leads to the
factorized expression,

\bq\label{2.7}
{1\over{N}} {\partial\over{\partial \lambda}}\rho(\lambda) = -
\phi(\lambda)\psi(\lambda)
\eq
where
\bq\label{2.8}
\phi(\lambda) = \int_{-\infty}^{\infty} {dt\over{2\pi}} e^{-{N\over{2}} t^2
+ {N\over{2}}
\ln (a^2 + t^2) - N i t \lambda}
\eq
\bq\label{2.9}
\psi(\lambda) = \oint {du\over{2 \pi i}} e^{-{N\over{2}}u^2 -
{N\over{2}}\ln( a^2 - u^2) + N u \lambda} \eq For $N$ large, we may apply
to the two integrals defining the functions $\phi$ and $\psi$ the
saddle-point method. When
$\lambda_1$ and $\lambda_2$ are near the origin the saddle-points in the
variables $t$ and $u$
move to the origin. Therefore for obtaining the large $N$ behavior of
$\phi$ near $\lambda =0$ we
can expand the logarithmic term in powers of $t$. One sees readily that the
coefficient of the quadratic term in $t^2$ vanishes for $a = 1$; in fact
three saddle-points are merging at the origin when $a$ reaches one. This is
the critical value at which the gap closes. We must then expand in the
exponential up to order $t^4$ and we obtain \bq\label{2.10} \phi(\lambda) =
\int_{-\infty}^{\infty} {dt\over{2 \pi}} e^{- {N\over{4}} t^4 - N i t
\lambda}
\eq
Rescaling $t$ to $N^{-1/4} t^{\prime}$ and setting $\lambda = N^{- 3/4}x$
we find that
\bq\label{2.11}
\hat\phi(x) = N^{1/4}\phi(N^{-3/4}x)
\eq
has a large $N$, finite $x$, limit given by

\bq\label{2.12}
\hat\phi(x) = {1\over{\pi}}\int_{0}^{\infty} dt e^{- {1\over{4}} t^4} \cos
( t x)
\eq
It is immediate to verify that it satisfies the differential equation,
\bq\label{2.13}
\hat\phi^{\prime\prime\prime}(x) = x \hat\phi(x) \eq
From the integral representation (\ref{2.8}) we obtain easily the Taylor
expansion of this function at the origin \bq\label{2.14} \hat\phi(x) =
{\sqrt{2}\over{4 \pi}}\sum_{m=0}^{\infty} {\Gamma({1\over{4}}
+{m\over{2}})(-1)^{m} 2^{m} x^{2 m}\over{(2 m)!}} \eq and its asymptotic
behavior at large x,
\bq\label{2.15}
\hat \phi(x) \sim \sqrt{{2\over{3 \pi}}} x^{-{1 \over{3}}} e^{-
{3\over{8}}x^{4\over{3}} }\cos( {3\sqrt{3}\over{8}}x^{4\over{3}} -
{\pi\over{6}})
\eq
For the second function (\ref{2.9}),
in the scaling limit, N large,
$\lambda$ small, $N^{3/4} \lambda$ finite, we may expand up to order $u^4$,
and define
\bq\label{2.16}
\hat\psi(x) = N^{1/4}\psi(N^{-3/4}x) .
\eq
In the large$N$, finite $x$, limit we find \bq\label{2.17} \hat\psi(x) =
\int_c {du\over{2\pi i}} e^{{u^4\over{4}} + u x}. \eq The integral is over
a path
consisting of four lines of
steepest descent in the complex u-plane. Along these straight lines, the
variable
$u$ is changed successively into
$e^{\pm {\pi\over{4}}i} u$ and
$e^{\pm {3
\pi\over{4}} i}u$. This leads to
\bq\label{2.18}
\hat\psi(x) = -{\rm Im} [ {\omega\over{\pi}} \int_{0}^{\infty}
e^{-{u^4\over{4}}}(
e^{x u \omega} - e^{-x u \omega} )]
\eq
in which $\omega = e^{{\pi i\over{4}}}$. The function $\hat\psi(x)$
satisfies the differential equation, \bq\label{2.19}
\hat\psi^{\prime\prime\prime}(x) = - x \hat\psi(x) \eq and again we find
from (\ref{3.12}) the Taylor expansion \bq\label{2.20} \hat\psi(x) = -
{1\over{\sqrt{\pi}}}
\sum_{n=0}^{\infty}{(-1)^n x^{4n + 1} (2n)!\over{n!(4n+1)!}} \eq and the
large x behavior
\bq\label{2.21}
\hat \psi(x) \sim 2 \sqrt{{2\over{3 \pi}}}x^{-{1 \over{3}}}
e^{{3\over{8}}x^{4\over{3}} } \cos( {3\sqrt{3}\over{8}}x^{4\over{3}} + {2
\pi\over{3}})
\eq

As shown in \cite{BH4}, one may express the whole kernel $K_N(\lambda_1,
\lambda_2)$ of (\ref{3.1}) in terms of the two functions $\hat \phi$ and
$\hat \psi$ in the scaling limit. Defining \bq\label{2.22}
\lambda_1 = N^{-3/4}x, \lambda_2
=N^{-3/4}y
\eq
\bq\label{2.23}
K(x,y) = N^{1/4}K_{N}(N^{3/4}\lambda_1,N^{3/4}\lambda_2) . \eq in the large
N, finite x and y, limit, we had shown that \bq\label{2.26} K(x,y) = {\hat
\phi^{\prime}(x)\hat\psi^{\prime}(y) - \hat\phi^{\prime\prime}(x) \hat\psi
(y) - \hat\phi(x)\hat\psi^{\prime\prime}(y)\over{x - y}}. \eq Note that if
follows from (\ref{2.12}) that $\phi(x)$ is an even function,
whereas $\psi(x)$ is odd (\ref{2.18}). It implies in particular
\bq\label{2.30} K(-x,-y) = K(x,y). \eq
Therefore the density of state
is given by
\bq\label{2.27}
\hat\rho(x) = - [\hat\phi^{\prime}(x)\hat\psi^{\prime\prime}(x) -
\hat\phi^{\prime\prime}(x) \hat\psi^{\prime}(x) + x \hat\phi(x)\hat\psi(x)]
\eq
In the large x limit, it behaves as $\rho(x)\sim x^{1/3}$. Hereafter, we
denote simply $\hat \phi$ and $\hat \psi$ by $\phi$ and $\psi$,
respectively.

\sect{ Fredholm theory
}

The level spacing function $E(s)$, the probability that there is no
eigenvalue inside the interval (-s/2, s/2) centered around the singular
point $s=0$,
is given by the Fredholm determinant,
\begin{eqnarray}\label{3.1}
E(a,b) &=& {\rm det}[ 1 - \hat K ]\nonumber\\ &=& \sum_{n=0}^{\infty}
{(-1)^{n}\over{n!}} \int_a^b \cdots \int_a^b \Pi_{k=1}^{n} d x_k {\rm det}[
K(x_i,x_j)]_{i,j=1,...,n} \end{eqnarray} if we choose for (a,b) the
interval (-s/2, s/2). The sine and Airy
kernels are symmetric kernels, $K(x,y) = K(y,x)$, whereas our kernel in
(\ref{2.26}) is not symmetric, since the two functions $\phi(x)$ and
$\psi(x)$ are different.Our notations here are as follows : the kernel
$K(x,y)$ is defined by (\ref{2.26}). However it acts on the interval (a,b)
; therefore we have used in (\ref{3.1}) the kernel $\hat K (x,y)$ defined
by the restriction of $K$ to the interval : \bq \label{3.50} \hat K(x,y) =
K(x,y) \theta (y-a) \theta (b-y) = K(x,y) \Theta(y)\eq
in which $\theta(x)$ is the Heaviside function, and we have used for
convenience the notation
\bq \label{3.51} \Theta(y) = \theta (y-a) \theta (b-y). \eq

In order to calculate the derivative of the logarithm of $E(a,b)$ with
respect to the end points
one writes
\bq
\ln E(a,b)= \rm{Tr}\ln(1-\hat K), \eq
and thus
\bq
{\partial {\ln}E(a,b)\over{\partial b}} = - \rm{Tr} {1\over{1-\hat
K}}{\partial \hat K \over{\partial b}}. \eq
>From (\ref{3.50}) we find
\bq {\partial \hat K(x,y)\over{\partial b}} = K(x,b) \delta(y-b), \eq and
therefore, if we introduce the
Fredholm resolvent $\tilde K(b,b)$ , defined by \bq \tilde K = {\hat
K\over{1 - \hat K}}
\eq
we obtain
\bq
{\partial {\ln}E(a,b)\over{\partial b}} = - \tilde K(b,b) \eq Similarly the
derivative with respect to $a$ is \bq
{\partial {\ln}E(a,b)\over{\partial a}} = \tilde K(a,a) \eq Therefore, when
we choose for (a,b) the interval (-s/2,s/2), we have \bq {d \ln E(s)\over{d
s}} = {1\over{2}}({\partial \over{\partial b}} - {\partial \over{\partial
a}})\ln E(s)|_{b=-a=s/2} = - \tilde K({s\over{2}},{s\over{2}})
\eq
This leads to
\bq\label{3.a8}
E(s) = \exp [ - \int_0^{s} \tilde K({s^{\prime}\over{2}},
{s^{\prime}\over{2}}) ds^{\prime}] \eq

We now define six functions, obtained by acting with the operator $(1 -
\hat K)^{-1}$ on the functions $\phi(x)$, $\psi(x)$ and their first two
derivatives,
\begin{eqnarray}\label{3.2} &&q_0(x) = ( 1 - \hat K )^{-1} \phi(x) \\ &=&
\phi(x) + \int_a^b K(x,y) \phi(y) dy + \int_a^b\int_a^b K(x,y) K(y,z)
\phi(z) dy
dz + \cdots \nonumber\end{eqnarray} We find it easier to use Dirac's
notations \bq\label{3.3}
q_0(b,a;x) = <x|{1\over{1 - \hat K}}|\phi>. \eq

Similarly we define
\bq\label{3.6}
q_n(b,a;x) = <x|{1\over{1 - \hat K}}|\phi^{(n)}> \eq where $\phi^{(n)}(x)$
is the n-th derivative of $\phi(x)$. For the function $\psi(x)$, we act
with the operator $( 1
- \hat K)^{-1}$ on bras, i.e. dual vectors, rather than on kets. Because of
the lack of symmetry of the kernel $K$ this
introduces a kernel $ \hat L$, and \bq\label{3.7} p_n(b,a;x) =
(-1)^{n-1}<\psi^{(2-n)}|{1\over{1 - \hat L}}|x> \eq where \bq\label{3.8}
\hat L(y,x) = \Theta(y) K(y,x)
\eq

When we set $x = b$, and $a = -b$, then the six functions $q_n$ and $p_n$
become functions of the single variable $b$, which we denote as
\bq\label{3.9}
Q_n(b) = q_n(b,-b;b), \qquad
P_n(b) = p_n(b,-b;b)
\eq

The calculation of the derivatives of these functions $Q_n(b)$ and $P_n(b)$
implies to consider separate variations in the functions $q_n(b,a;x)$ and
$p_n(b,a;x)$ with respect to $b$, $a$, and $x$, before we set $a=-b$ and
$x=b$. The calculation is tedious, but not difficult,
and we have given more details in the
Appendix C. The resulting differential equations are
\begin{eqnarray}\label{3.10}
\dot{Q_0} &=& Q_1 + {2\over{b}}Q_1 P_1 Q_0 \nonumber\\ \dot{Q_1} &=& Q_2 -
{2\over{b}} Q^2_1 P_1 - Q_0 u \nonumber\\ \dot{Q_2} &=& b Q_0 +
{2\over{b}} Q_1 P_1 Q_2 - Q_1 v \\
\dot{P_0} &=& - b P_2 - {2\over{b}} Q_1 P_1 P_0 + P_1 u
\nonumber\\
\dot{P_1} &=& - P_0 + {2\over{b}} Q_1 P^2_1 + P_2 v \nonumber\\
\dot{P_2} &=&- P_1 - {2\over{b}} Q_1 P_1 P_2
\nonumber
\end{eqnarray}
where a dot means taking the derivative with respect to $b$. The two
auxiliary functions $u$ and $v$ are defined as \bq\label{3.12}
u = < \psi|q_1>, \qquad
v = <\psi^{\prime}\vert q_0>
\eq

and they satisfy
\bq\label{3.14}
\dot{u} = - 2 P_2 Q_1 , \qquad
\dot{v} = 2 P_1 Q_0.
\eq
From these equations, we find succesively the relations
\bq\label{3.15}
u + v = - 2 P_2 Q_0,
\eq
then \bq\label{3.16} \dot{Q_0} = Q_1 ( 1 + {\dot{v}\over{b}}), \qquad
\dot{P_2} = - P_1 ( 1 - {\dot{u}\over{b}}) \eq Using (\ref{3.14}) and
(\ref{3.15}),
we obtain from the second equation of (\ref{3.10}), \bq\label{3.18} - 2 P_2 Q_2
= \ddot{u} - {\dot{u}\dot{v}\over{u + v}} + {2\over{b}}
{\dot{v}\dot{u}^2\over{u + v}} + u ( u + v) \eq and using the fifth
equation of (\ref{3.10}), we have
\bq\label{3.19} - 2 Q_0P_0 = \ddot{v} - {\dot{u}\dot{v}\over{u + v}} -
{2\over{b}} {\dot{u}\dot{v}^2\over{u + v}} + v ( u + v). \eq Taking the
derivatives of these two
equations and using the third and sixth equations of (\ref{3.10}), we end
up with two coupled non-linear equations \bq\label{3.20} {d^3 u\over{d
b^3}} +
({2\dot{u}\over{b}} - 1) [ b( u + v) + {1\over{u + v}}( \ddot{v}\dot{u} + 2
\dot{v}\ddot{u}) - {\dot{u}\dot{v}\over{( u + v)^2}}( 2 \dot{v} + \dot{u})]
- {2
\dot{v}(\dot{u})^2\over{b^2(u + v)}} = 0 \eq
\bq\label{3.21}
{d^3 v\over{d b^3}} + ({2\dot{v}\over{b}} + 1) [ b( u + v) - {1\over{u +
v}}( \ddot{u}\dot{v} + 2 \dot{u}\ddot{v}) + {\dot{u}\dot{v}\over{( u +
v)^2}}( 2 \dot{u} + \dot{v})] + {2 \dot{u}(\dot{v})^2\over{b^2(u + v)}} = 0
\eq
In the large $b$ limit, the asymptotic behavior of the solutions of these
equations is obtained under the form
\bq\label{3.22}
u = {b^2\over{4}} + {A\over{2}} b^{2/3} + \cdots \eq \bq\label{3.23} v = -
{b^2\over{4}} + {A\over{2}} b^{2/3} + \cdots \eq Inserting these
expressions in (\ref{3.20}), we find $A$ from the coefficient of the terms
of order $b^{1/3}$,
\bq\label{3.24}
A = - ({1\over{4}})^{1/3}
\eq

The kernel $\tilde K(b,b)$ may then expressed as
\begin{eqnarray}\label{3.25} \tilde
K(b,b) &=& b P_2 Q_0 + Q_2 P_1 + Q_1 P_0 - u P_1 Q_0 - v P_2 Q_1 \nonumber\\
&-&
{1\over{2b}}(P_1^2
Q_1^2 - Q_2 P_2 - Q_0 P_0)^2
\end{eqnarray}
Remarkably this kernel at coinciding points is in fact a Hamiltonian, that
we denote as
\bq\label{3.100} H(b) = \tilde K(b,b),\eq from which Hamilton's equations
\bq\label{3.500}
\dot {Q}_n = {\partial H\over{\partial P_n}}, \qquad \dot{P}_n = -
{\partial H\over{\partial Q_n}} \eq
coincide with the differential equations (\ref{3.10}). This allows one to
obtain a simple expression for
its derivative with respect to $b$ becomes \begin{eqnarray}\label{3.600} {d
H(b)\over{d b}} &=& {\partial H\over{\partial b}}\nonumber\\ &=& Q_0 P_2 +
{2\over{b^2}} P_1^2 Q_1^2 \end{eqnarray}
In terms of
$u$ and $v$, it becomes
\bq
{d H(b)\over{d b}} = - {u + v\over{2}} + {(\dot{u}\dot{v})^2\over{ 2 b^2(u
+ v)^2}}
\eq
Thus, from the previous result, we get the large $b$ behavior \bq {d
H(b)\over{d b}} \sim 5\cdot 2^{-11/3} b^{2/3} \eq. Integrating once we find
for the hamiltonian
$H(b) \sim 3\cdot 2^{-11/3} b^{5/3} \sim 3\cdot 2^{-16/3} s^{5/3}$ in the
large $s$ limit.

From (\ref{3.a8}), we have
\begin{eqnarray}\label{3.160}
E(s) &=& \exp[ - \int_0^s H(s^{\prime}) ds^{\prime}]\nonumber\\ &\sim&
\exp[ - 9\cdot 2^{-25/3} s^{8/3} ] \end{eqnarray}

We have thus derived the exponent $s^{8/3}$ which was a mere conjecture in
our previous
article \cite{BH4}. There we had performed a
simple Pad\'e analysis of the small $s$ expansion, and assumed that it was
Gaussian at large $s$ in
the variable $\tilde s$. This had led us to the estimate \bq E(s) \sim
\exp[ - 0.332 \tilde s^2],
\eq
or since $\tilde s \sim {3\sqrt{3}\over{2 \pi}} ({1\over{2}})^{4/3} s^{4/3}$,
$E(s) \sim \exp[ - 0.0358 s^{8/3}]$. Our analytic result (\ref{3.16}) gives
$E(s)\sim \exp[ - 0.0280 s^{8/3}]$. Thus the estimation by a simple Pad\'e
analysis was not too far from the exact result.

\sect{SUMMARY AND DISCUSSION }

In this paper, we
have investigated the level spacing probability for the case, in which two
edge singularities collapse. By the use of Fredholm theory, we have derived
an expression for the level spacing probability, whose logarithmic
derivative turns out
to act
as a Hamiltonian. The same strategy allows one to solve as well the simpler
cases, such as the sine-kernel (relevant to the level spacing for ordinary
non-singular)
points of the spectrum and the Airy kernel (which applies to a single edge
singularity).
The corresponding Hamilton's equations determine fully the level spacing,
and in particular one can obtain analytically its asymptotic expansion at
large spacing $s$. This allowed us
to confirm the conjecture that we had
made in our previous article, on the asymptotic Gaussian behavior of the
level spacing in terms
of the variable $\tilde s = \int_{-s/2}^{s/2}\rho(x) dx$, which is the
number of eigenvalues in the interval of size $s$ \cite{BH4}.
 We have thus derived here that the level spacing probability $E(s)$
behaves $\exp[ - C s^{{8\over{3}}}]$, with a constant $C$ that we have
analytically
determined. More generally  the three cases that we have solved are
consistent with the asymptotic Gaussian behavior of
 $E(s)$ with respect to $\tilde s$, i.e. to
$E(s)\sim\exp[ - C s^{2 \beta + 2}]$ for large spacing $s$, whenever the
interval of size $s$
is around a point at which the density of state behaves
as $\rho(\lambda)\sim \lambda^{\beta}$. The scaling variable $s$ is in fact
related to the actual interval between the eigenvalues by absorbing a
power $N^{1/(\beta+1)}$. If we let $N$ go to infinity first, at fixed
interval, in which case $s$ is also large, a finite limit of ${\ln
E(s)\over{N^2}}$
for large $N$,
implies that $E(s)$ does fall for large $s$ as $E(s)\sim\exp[ - C s^{2
\beta + 2}]$. In other words, $E(s)$ behaves for large$N$, fixed interval,
as a partition
function.

\begin{center}
{\bf Acknowledgement}
\end{center}
It is a pleasure to thank Prof. Leonid Pastur for stimulating discussions.
This work was supported by the CREST of JST. S. H. thanks a Grant-in-Aid
for Scientific Research by the Ministry of Education, Science and Culture.

\newpage
\setcounter{equation}{0}
\renewcommand{\theequation}{A.\arabic
{equation}}
{\bf Appendix A: {Level spacing probability for the sine kernel}} \vskip 5mm

The sine kernel is defined by
\bq\label{A.1}
K(x,y) = {\phi(x)\phi^{\prime}(y) - \phi(y)\phi^{\prime}(x) \over{x - y}}
\eq where $\phi(x) = \sin x$ satifies $\phi^{\prime\prime} (x) = -
\phi(x)$. (For convenience we have absorbed in the normalization the usual
factor $\pi$). We
consider $q(x)$ and $p(x)$ defined by
\bq\label{A.2}
q(x) = <x| {1\over{1 - \hat K}}|\phi>, \qquad p(x) =
<\phi^{\prime}|{1\over{1 - \hat K}}|x> \eq and the Fredholm resolvent
$\tilde K$ \bq\label{A.3}
\tilde K = {\hat K\over{1 - \hat K}}
\eq
We have
\begin{eqnarray}\label{A.4}
(x - y) \tilde K &=& <x|[X,\tilde K]|y> = <x|[X,{1\over{1 - \hat
K}}]|y>\nonumber\\ &=& <x| {1\over{1 - \hat K}}[X,\hat K]{1\over{1 - \hat
K}}|y> \end{eqnarray}
The definition (\ref{A.1}) of the kernel reads \bq\label{A.5}
[X,K] = |\phi><\phi^{\prime}| - |\phi^{\prime}><\phi| \eq since \bq\label{A.6}
(x - y) K(x,y) = <x|[X,K]|y>
\eq
Thus, from (\ref{A.4}), we obtain
\bq\label{A.7}
\tilde K(x,y) = {q(x) p(y) - q(y) p(x)\over{x - y}} \eq

Since the functions $q(x)$ and $p(x)$ depend upon the interval (a,b), we
denote them more precisely as $q(b,a;x)$ and $p(b,a;x)$. We then set $x=b$
and vary $b$, i.e. take the derivative of $ q(b,a;b)$ at fixed $a$
\bq\label{A.8}
{\partial q(b,a;b)\over{\partial b}} = <b|D{1\over{1 - \hat K}}|\phi> +
<b|{1\over{1 - \hat K}}{\partial \hat K\over{\partial b}}{1\over{1 - \hat
K}}|\phi> \eq
where $D$ is the derivative operator: $<x\vert D\vert f>= f'(x)$. From the
definition of $\hat K$ we have,
\bq\label{A.9}
{\partial \hat K(x,y)\over{\partial b}} = K(x,y) \delta(b - y) = K|b><b|
\eq Thus the second term of (\ref{A.8}) becomes $\tilde K(b,b) q(b)$. The
first term of (\ref{A.8}) becomes
\begin{eqnarray}\label{A.10}
<b|D{1\over{1 - \hat K}}|\phi>
&=& p(b) + <b|[D,{1\over{1 - \hat K}}]|\phi>\nonumber\\ &=& p(b) +
<b|{1\over{1 - \hat K}}[D,\hat K]{1\over{1 - \hat K}}|\phi> \end{eqnarray}
Since
\begin{eqnarray}\label{A.11}
<x|[D,\hat K]|y> &=& ({\partial\over{\partial x}} + {\partial\over{\partial
y}})\hat K(x,y)\nonumber\\ &=& K(x,y) (\delta(y - a) - \delta(y - b))
\end{eqnarray}
we obtain
\bq\label{A.12}
[D,\hat K] = K|a><a| - K|b><b|
\eq
>From (\ref{A.8}) and (\ref{A.10}), we have
\bq\label{A.13}
{\partial q(b,a;b)\over{\partial b}} = p(b,a;b) + \tilde K(b,a) q(b,a;a)
\eq Similarly, we have for $p(b)$ as
\bq\label{A.14}
{\partial p(b,a;b)\over{\partial b}} = - q(b,a;b) + \tilde K(b,a) p(b,a;a)
\eq We denote $q(b,-b;b)$ by $Q(b)$,
\bq\label{A.15}
{d Q(b)\over{d b}} = {\partial q(b,a;b)\over{\partial b}}|_{a = -b} -
{\partial q(b,a;b)\over{\partial a}}|_{a = -b} \eq Since
\begin{eqnarray}\label{A.16}
{\partial q(b,a;b)\over{\partial a}} &=& <b|{1\over{1 - \hat K}}({\partial
\hat K\over{\partial a}}){1\over{ 1 - \hat K}}|\phi>\nonumber\\ &=& -
\tilde K(b,a)q(b,a;a)
\end{eqnarray}
we have
\bq\label{A.17}
\dot{Q}(b) = P(b) + 2 \tilde K(b,-b)Q(-b) \eq where \bq\label{A.18}
\tilde K(b,-b) = {Q(b) P(b)\over{b}}
\eq
Note that $Q(-b) = - Q(b)$ and $P(-b)= P(b)$. Similarly we have an equation
for $P(b)$. Thus we obtain finally
\begin{eqnarray}\label{A.19}
\dot Q &=& P( 1 - {2Q^2\over{b}})\nonumber\\ \dot P &=& Q( {2 P^2\over{b}}
- 1) \end{eqnarray}

The function $\tilde K(b,b)$ is related to $P$ and $Q$ as \bq\label{A.20}
\tilde K(b,b) = P^2 + Q^2 - {2 P^2 Q^2\over{b}} \eq which gives the
logartihmic derivative of the level spacing probability $E(s)$. Noting $b =
s/2$, we have
\begin{eqnarray}\label{A.21}
{d Q\over{d s}} &=& {P\over{2}}( 1 - {4 \over{s}} Q^2)\nonumber\\ {d
P\over{d s}} &=& {Q\over{2}} ( {4 P^2\over{s}} - 1) \end{eqnarray} In the
large $s$ limit,
we have $Q \sim P \sim s^{1/2}/2$. From (\ref{A.20}), we obtain $H(s) =
\tilde K(b,b) \sim s/4$.
The level spacing probability $E(s)$ behaves thus in the large $s$ limit as
\begin{eqnarray}
E(s) &\sim& \exp[ - \int_0^{\pi s} {s^{\prime}\over{4}} ds^{\prime}
]\nonumber\\ &\sim& \exp[ - {\pi^2\over{8}} s^2 ] \end{eqnarray}
For small s, we have by solving (\ref{A.21}) iteratively, with the initial
condition $P(0) = 1$, $Q(0)=0$ , \bq Q = {s\over{2}} - {s^3\over{48}} +
O(s^5), \qquad P = 1 + s + {7\over{8}}s^2 + O(s^3) \eq
Then from (\ref{A.20}), $H(b) = \tilde K(b,b)$ behaves for small $s$ as \bq
H(s) = 1 + s + s^2 + O(s^3)
\eq
This leads to
\begin{eqnarray}
E(s) &=& \exp[-\int_0^s H(s^{\prime}) ds^{\prime}]\nonumber\\ &=& 1 - s +
O(s^4)
\end{eqnarray}
which is consistent with all the well-known results on this well-studied
case \cite{Mehta}.

\vskip 5mm
\setcounter{equation}{0}
\renewcommand{\theequation}{B.\arabic
{equation}}
{\bf Appendix B: {Level spacing probability for the Airy kernel}} \vskip 5mm

The sine kernel applies to a regular point of the spectrum. However when
one studies the vicinity of the edge of the spectrum, in the appropriate
scaling limit, the correlation functions and the level spacing are given by
an Airy kernel. The
level spacing has been studied by Tracy and Widom \cite{Tracy1}. We repeat
here the same technique. Consider the interval $(-s/2,s/2)$
as in the sine case.
We denote the Airy function $A_i(x)$ by $\phi(x)$, it satisfies
$\phi^{\prime\prime}(x) = x \phi(x)$.
We use the same notation $q(x)$ and $p(x)$ as in (\ref{A.2}). As for the
sine case, the Fredholm resolvent $\tilde K(a,b)$ is given by \bq
\tilde K(a,b) = {q(a)p(b) - p(a)q(b)\over{a - b}} \eq
We have
\bq
({\partial \over{\partial x}} + {\partial \over{\partial y}}) K(x,y ) = -
\phi(x)
\phi(y)
\eq
Therefore, we have
\bq
[D,K] = - |\phi><\phi|\Theta + K|a><a| - K|b><b| \eq
Thus, by the same procedure as sine case, we get \bq
{\partial q(b)\over{\partial b}} = p(b) - q u + \tilde K(b,a) q(a) \eq
\bq
{\partial p\over{\partial b}} = b q(b) + u p(b) - 2 q(b) v + \tilde K(b,a)
p(a) \eq
where
\bq
u = <\phi|q>,\qquad
v = <\phi|p>
\eq
The Fredholm resolvent $\tilde K(b,b)$ becomes \begin{eqnarray}
\tilde K(b,b) = p^2(b) - b q^2(b) - 2 u p(b) q(b) + 2 q^2(b) v \nonumber\\
+ {1\over{b - a}}[q(b)p(a) - p(b)q(a)][q(a)p(b) - p(a)q(b)] \end{eqnarray}
Again $H(b) =\tilde K(b,b)$ acts as a Hamiltonian, since we have \bq
{\partial H(b)\over{\partial p(b)}} = 2 {\partial q(b)\over{\partial b}},
\qquad
{\partial H(b)\over{\partial q(b)}} = - 2 {\partial p(b)\over{\partial b}} \eq
The derivative of the Hamiltonian becomes \bq
{d H(b)\over{d b}} = - q^2(b) + {(q(b) p(a) - p(b) q(a))^2\over{(b - a)^2}} \eq
In the Airy kernel, due to the parity around the edge, $H(a)$ and $H(b)$
are different. The quantity $q(b)$ and $p(b)$ becomes exponentially small
in the large $b$ limit ($b > 0$) as same as Airy function. We have for $a =
- s/2$,
\bq
{\partial q(a)\over{\partial a}} = p(a) - q(a) u - \tilde K(a,b) q(b) \eq
\bq
{\partial p(a)\over{\partial a}} = a q(a) + u p(a) - 2 q(a) v - \tilde
K(a,b) p(b)
\eq
The Hamiltonian $H(a)$ becomes
\begin{eqnarray}
H(a) &=& p^2(a) - a q^2(a) - 2 u p(a) q(a) + 2 q^2(a) v \nonumber\\
&+& \tilde K(a,b)[q(a)p(b) - p(a)q(b)]
\end{eqnarray}
Since $\tilde K(a,b)$ can be neglected for the large $b$ limit, and we have
a relation,
$u^2 - 2 v = q^2$, we get
\bq
{d^2 q(a)\over{d a^2}} = a q(a) + 2 q^3(a) \eq

This leads to $q(a) \sim \sqrt{s}/2$, and

$H(a)
\sim s^2/16$ in the large s limit.
Then, we get
$E(s) \sim \exp[ - s^3/96]$.
In order to check the usefulness of the Pad\'e analysis we expand now at
small s.

The Airy function $A(x)$ has the Taylor expansion for small x, \begin{eqnarray}
A(x) &=& c_1[ 1 + {1\over{6}} x^3 + {4\over{6!}} x^6 + \cdots]\nonumber\\
&-& c_2 [ x + {2\over{4!}} x^4 + {2 \cdot 5\over{7!}}x^7 + \cdots]
\end{eqnarray}
where $c_1 = 3^{-2/3}/\Gamma(2/3)$ and $c_2 = 3^{-1/3}/\Gamma(1/3)$. If we
use $\tilde s$ , defined in (\ref{1.5}), it is related to $s$ in this Airy
case by
\bq
\tilde s = c^2_2 s + {c_1 c_2\over{12}} s^3 - {c^2_1\over{960}} s^5 -
{c^2_2\over{16128}}
s^7 + O(s^9)
\eq
The level spacing $E(s)$ is thus expanded in powers of $\tilde s$ as
\begin{eqnarray}
E(s) &=& 1 - \tilde s + {1\over{6}}({c_1 c^3_2\over{3}} - {c^4_1\over{8}})
{1\over{c^8_2}} \tilde s^4
+ \cdots
\nonumber\\
&=& 1 - \tilde s + 0.544868 \tilde s^4 - 42.5418 \tilde s^6 + O(\tilde s^8)
\end{eqnarray}
We now apply a Pad\'e analysis to $H(s)$, or rather to \begin{eqnarray}
R(\tilde s) &=& {d \over{d \tilde s}} {\ln} E(\tilde s)\nonumber\\ &=& - {1
+ a_1 \tilde s + a_2 \tilde s^2 + a_3 \tilde s^3\over{ 1 + b_1 \tilde s +
b_2 \tilde s}}
\end{eqnarray}
where we have $a_1 = 754.156$, $a_2=1640.76$, $a_3=1638.58$, $b_1 =
753.156$, and $
b_2 = 886.601$. From $E(\tilde s)$, differentiating twice with respect to
$\tilde s$ ,
we obtain $p(\tilde s)$,
which is slightly different from the usual 'Wigner surmise' function of the
sine case.

\vskip 5mm
\setcounter{equation}{0}
\renewcommand{\theequation}{C.\arabic
{equation}}
{\bf Appendix C: { Level spacing probability for the gap closure kernel }}
\vskip 5mm

We consider now the kernel
\bq\label{C.1}
K(x,y) = {\phi^{\prime}(x)\psi^{\prime}(y) - \phi^{\prime\prime}(x)\psi(y)
- \phi(x)\psi^{\prime\prime}(y)\over{x - y}}, \eq or, in operator
notations, \bq\label{C.2}
[X,K] = |\phi^{\prime}><\psi^{\prime}| - |\phi^{\prime\prime}><\psi| -
|\phi><\psi^{\prime\prime}| \eq
Similarly to (\ref{A.4}), we have
\begin{eqnarray}\label{C.3}
(x - y) \tilde K
&=& <x| {1\over{1 - \hat K}}[X,\hat K]{1\over{1 - \hat K}}|y>\nonumber\\
&=& q_1(x)p_1(y) - q_2(x)p_0(y) - q_0(x)p_2(y) \end{eqnarray} The
derivative of $q_n(b,a;b)$ for a fixed $a$ becomes
\begin{eqnarray}\label{C.4}
{\partial q_n(b,a;b) \over{\partial b}} &=& <b|D{1\over{1 - \hat
K}}|\phi^{(n)}> + <b|{1\over{1 - \hat K}}({\partial \hat K\over{\partial
b}}){1\over{1 - \hat K}}|\phi^{(n)}> \nonumber\\ &=& q_{n+1}(b,a;b) +
<b|{1\over{1 - \hat K}}[D,\hat K]{1\over{1 - \hat K}}|\phi^{(n)}>
\nonumber\\
&& + <b|{\hat k\over{1 - \hat K}}|b><b|{1\over{1 - \hat K}}|\phi^{(n)}>
\end{eqnarray}
We have also
\begin{eqnarray}\label{C.5}
<x|[D,\hat K]|y> &=& ({\partial\over{\partial x}} + {\partial\over{\partial
y}}) K(x,y) + <x|K|a><a|y> \nonumber\\
&-& <x|K|b><b|y>
\end{eqnarray}
The first term is simply $- \phi(x)\psi(y)$. This leads to \bq\label{C.6}
[D,\hat K] = - |\phi><\psi|\Theta + K|a><a| - K|b><b| \eq Therefore
\bq\label{C.7}
{\partial q_n(b,a;b)\over{\partial b}}= q_{n+1} + \tilde K(b,a)q_n(a) -
q_0(b)<\psi|q_n> \eq
and
\begin{eqnarray}\label{C.8}
q_3 &=& <x|{1\over{1 - \hat K}} X|\phi> = <x|X{1\over{1 - \hat K}}|\phi> +
<x|[{1\over{1 - \hat K}},X]|\phi> \nonumber\\ &=& x q_0(x) + <x|{1\over{1 -
\hat K}}[\hat K,X]{1\over{1 - \hat K}}|\phi> \nonumber\\ &=& x q_0 - v_2
q_1 + u_1 q_2 + v_3 q_0
\end{eqnarray}
where $u_1 = <\psi|q_0>$, $v_2 = <\psi^{\prime}|q_0>$ and $v_3 =
<\psi^{\prime\prime}|q_0>$.

The function $p_n(x)$ is defined by
\bq\label{C.9}
p_n(x) = (-1)^{n-1}<\psi^{(2 - n)}|{1\over{1 - \hat L}}|x> \eq
where
\bq\label{C.10}
\hat L(y,x) = \Theta(y) K(y,x) \eq We have \bq\label{C.11}
[D,\hat L] = - \Theta|\phi><\psi| + |a><a|K - |b><b|K, \eq in which
$\Theta$ is a local operator defined by \bq
<y|\Theta|y^{\prime}> = \delta(y - y^{\prime})\theta(y - a) \theta(b - y)
\eq
Thus we obtain
\bq\label{C.12}
{\partial p_n(b)\over{\partial b}} = - p_{n-1}(b) - p_0(b)
<\psi^{(2-n)}|q_0> + p_n(a) \tilde K(a,b) \eq
with $p_{-1}(x)$ obtained as
\begin{eqnarray}\label{C.13}
p_{-1}(x) &=& - <\psi^{\prime\prime\prime}|{1\over{1 - \hat
L}}|x>\nonumber\\ &=& x p_2(x) - <\psi|{1\over{1 - \hat L}}[x,\hat
L]{1\over{1 - \hat L}}|x> \\
&=& - x p_0(x) - p_1(x)<\psi|q_1> - p_2 <\psi|q_2> - p_0(x) <\psi|q_0>
\nonumber
\end{eqnarray}
where
\begin{eqnarray}
<y|[X,\hat L]|x> &=& ( y - x) \Theta(y) K(y,x)\nonumber\\ &=&
\Theta(y)(|\phi^{\prime}><\psi^{\prime}| - |\phi^{\prime\prime}><\psi| -
|\phi><\psi^{\prime\prime}|) \end{eqnarray}

The function $\phi(x)$ is an even function of $x$, and $q_0(x)$ becomes an
even function. The function $\psi(x)$ is a odd function. Therefore, we have
$u_1 = v_3 =0$ for the interval (-b,b). Also we $<\psi|q_2> = <\psi|q_0> =
<\psi^{\prime\prime}|q_0> = 0$. Nonvanishing quantities are $u_2 =
<\psi|q_1>$, and $v_2 = <\psi^{\prime}|q_0>$. We denote them simply by $u$
and $v$.

Noting that
\bq
\dot Q_n = {\partial q_n\over{\partial b}}|_{a = -b} - {\partial
q_n\over{\partial b}}|_{a = - b} \eq
we obtain (\ref{3.10}). The equations for $\dot{P}_n$ in (\ref{3.10}) are
also obtained
similarly.

The derivative of $u = u_2$ becomes
\bq
\dot{u} = - 2 P_2(b) Q_1(b)
\eq
This is obtained as
\bq
{\partial u\over{\partial b}} = < \psi| \Theta {1\over{1 - \hat
K}}{\partial \hat K\over{\partial b}}{1\over{1 - \hat K}}|\phi^{\prime}> +
\psi(b)q_1(b) \eq
Using
\bq
{\partial \hat K\over{\partial b}} = K(x,y)\delta(y - b) = K|b><b| \eq we have
\bq
{\partial u\over{\partial b}} = - p_2(b) q_1(b) \eq The derivative of $u$
by $a$ becomes, similarly \bq {\partial u\over{\partial a}} = p_2(a)q_1(a)
\eq Putting $a = -b$, we have
\begin{eqnarray}
\dot{u} &=& {\partial u\over{\partial b}}|_{a = -b} - {\partial
u\over{\partial a}}|_{a = -b} \nonumber\\ &=& - 2 P_2(b) Q_1(b)
\end{eqnarray}
This is Eq.(\ref{3.12}).
\vskip5mm
\setcounter{equation}{0}
\renewcommand{\theequation}{D.\arabic
{equation}}
{\bf Appendix D: { Modified kernel }}
\vskip 5mm

We have considered the case of the fixed external source eigenvalues at $a
= \pm 1$.
Here, we take this external eigenvalue $a$ as $ a^2 = 1 + 2N^{-1/2}
\alpha$. The parameter $\alpha$ measures the approach to the limit $a = \pm
1$ in the large N limit, where the gap is closed. This change of $a$
modifies the function $\phi(x)$ and $\psi(x)$. We have in the large scaling
limit
\bq\label{4.1}
\phi(\lambda) = \int_{-\infty}^{\infty} {dt\over{2 \pi}} e^{-{t^4\over{4}}
- \alpha t^2 + i t \lambda} \eq
and it satisfies
\bq\label{4.2}
\phi^{\prime\prime\prime} - 2 \alpha \phi^{\prime} - \lambda \phi = 0 \eq
The equation, which $\psi(\lambda)$ satisfies, is also modified as
\bq\label{4.3}
\psi^{\prime\prime\prime} - 2 \alpha \psi^{\prime} + \lambda \psi = 0 \eq

Following the same procedure in the Appendix B in \cite{BH4}, we have a
kernel, which is slightly different from (\ref{3.6}), \bq\label{4.4} K(x,y)
= { \phi^{\prime}(x)\psi^{\prime}(y) - \phi^{\prime\prime}(x) \psi (y) -
\phi(x)\psi^{\prime\prime}(y) + 2\alpha \phi(x) \psi(y)\over{x - y}} \eq
However, the density of state $\rho(x)$ is expressed by the same equation
as (\ref{2.27}),
\bq\label{4.5}
\rho(x) = - [\phi^{\prime}(x)\psi^{\prime\prime}(x) -
\phi^{\prime\prime}(x) \psi^{\prime}(x)
+ x \phi(x)\psi(x)]
\eq

The large $\lambda$ behavior of $\phi(\lambda)$, for a fixed $\alpha$, is
obtained by a saddle point method. If we make a change of $t$ by $
\lambda^{1/3} t$, we find that the new term $\alpha t^2$ becomes negligible
compared with other terms, which becomes order of $\lambda^{4/3}$. Then, we
obtain the large $x$ behavior of $\rho(x)$ as $x^{1/3}$ same as before.

Using the same definitions for $q_n$ and $p_n$, we have an equation,
\bq\label{4.6} \tilde K(a,b) = {q_1(a)
p_1(b) + q_0(a) p_0(b) + q_2(a) p_2(b) - 2 \alpha q_0(a) p_2(b)\over{a -
b}} \eq

Although there are modifications in the differential equations for $q(b)$
and $p(b)$, the derivative of the Hamiltonian is given by
(\ref{3.600}), which gives the same asymptotic behavior of $E(s)$. We note
that the function $\phi(x)$ in (\ref{4.1}) appears for the second
Painlev\'e $A_2$ Garnier system (Appendix A in \cite{BH4}), and the
function $\phi(x)$ satisfies the coupled linear partial differential
equations about $x$ and $\alpha$.

\end{document}